\documentclass{article}
\usepackage{ijcai24}

\usepackage{times}
\usepackage{soul}
\usepackage{url}
\usepackage[hidelinks]{hyperref}
\usepackage[utf8]{inputenc}
\usepackage[small]{caption}
\usepackage{graphicx}
\usepackage{amsmath}
\usepackage{amsthm}
\usepackage{booktabs}
\usepackage{algorithm}
\usepackage{algorithmic}
\usepackage[switch]{lineno}

\graphicspath{ {images/} }

\title{Flexible Control in Symbolic Music Generation via Musical Metadata}

\author{
Sangjun Han
\and
Jiwon Ham\and
Chaeeun Lee\and
Heejin Kim\and
Soojong Do\and
Sihyuk Yi\and\\
Jun Seo\and
Seoyoon Kim\and
Yountae Jung\And
Woohyung Lim\\
\affiliations
LG AI Research\\
\emails
\{sj.han, jiwon.ham, chaen.lee, heejin.kim, dsj, sihyuk.yi,\\
jun.seo, seoyoon.kim, y.jung, w.lim\}@lgresearch.ai
}

\begin{document}

\maketitle

\begin{abstract}
    In this work, we introduce the demonstration of symbolic music generation, focusing on providing short musical motifs that serve as the central theme of the narrative.
    For the generation, we adopt an autoregressive model which takes musical metadata as inputs and generates 4 bars of multitrack MIDI sequences.
    During training, we randomly drop tokens from the musical metadata to guarantee flexible control.
    It provides users with the freedom to select input types while maintaining generative performance, enabling greater flexibility in music composition.
    We validate the effectiveness of the strategy through experiments in terms of model capacity, musical fidelity, diversity, and controllability.
    Additionally, we scale up the model and compare it with other music generation model through a subjective test.
    Our results indicate its superiority in both control and music quality.
    We provide a URL link\footnote{https://www.youtube.com/watch?v=-0drPrFJdMQ} to our demonstration video.
\end{abstract}

\section{Introduction}

Music composition requires an imaginative inspiration and a solid understanding of musical knowledge to convey artistic expression.
However, these components pose challenges for the public to achieve musical objectives.
To address this, we propose the development of a music interactive system that enables us to communicate with AI.
The desirable system should translate rough ideas into tangible forms of sound, allowing individuals to explore and express their musical identity.

The controllability of generative models is an important factor in that it allows creators to conduct the creative process.
In recent studies, various modalities such as text and video can serve as potential options for conditional inputs~\cite{tang2023any}.
While MIDI offers a defined protocol with standard instruments and note-timing information, there is a lack of well-established MIDI datasets for multimodal pairs.
To overcome this limitation, we leverage musical metadata as generative conditions, which can be extracted without any human effort.
This motivation is closely aligned with FIGARO~\cite{von2022figaro} and ComMU~\cite{lee2022commu}, but we extend this concept to large-scale MIDI.
While the previous works have required users to control all inputs, we enhance the generative model capability to adapt variable conditions to guarantee flexible controllability.

In this work, we introduce the demonstration of symbolic music generation, which generates 4 bars of multitrack MIDI sequences.
The interface is intuitive to follow; users can easily configure musical metadata for controllability.
With the help of our training strategy, it is not required for users to complete all conditions, allowing for default blank options.
We validate the effectiveness of the strategy in terms of model capacity, musical fidelity, diversity, and controllability.
Additionally, we scale up the model and conduct a human listening test to demonstrate its superiority over other music generation models.
Our demonstration contributes to providing musical motifs that serve as the central theme of the narrative.

\section{User Interface}

\begin{figure*}[!t]
  \centering
  \includegraphics[width=0.8\linewidth]{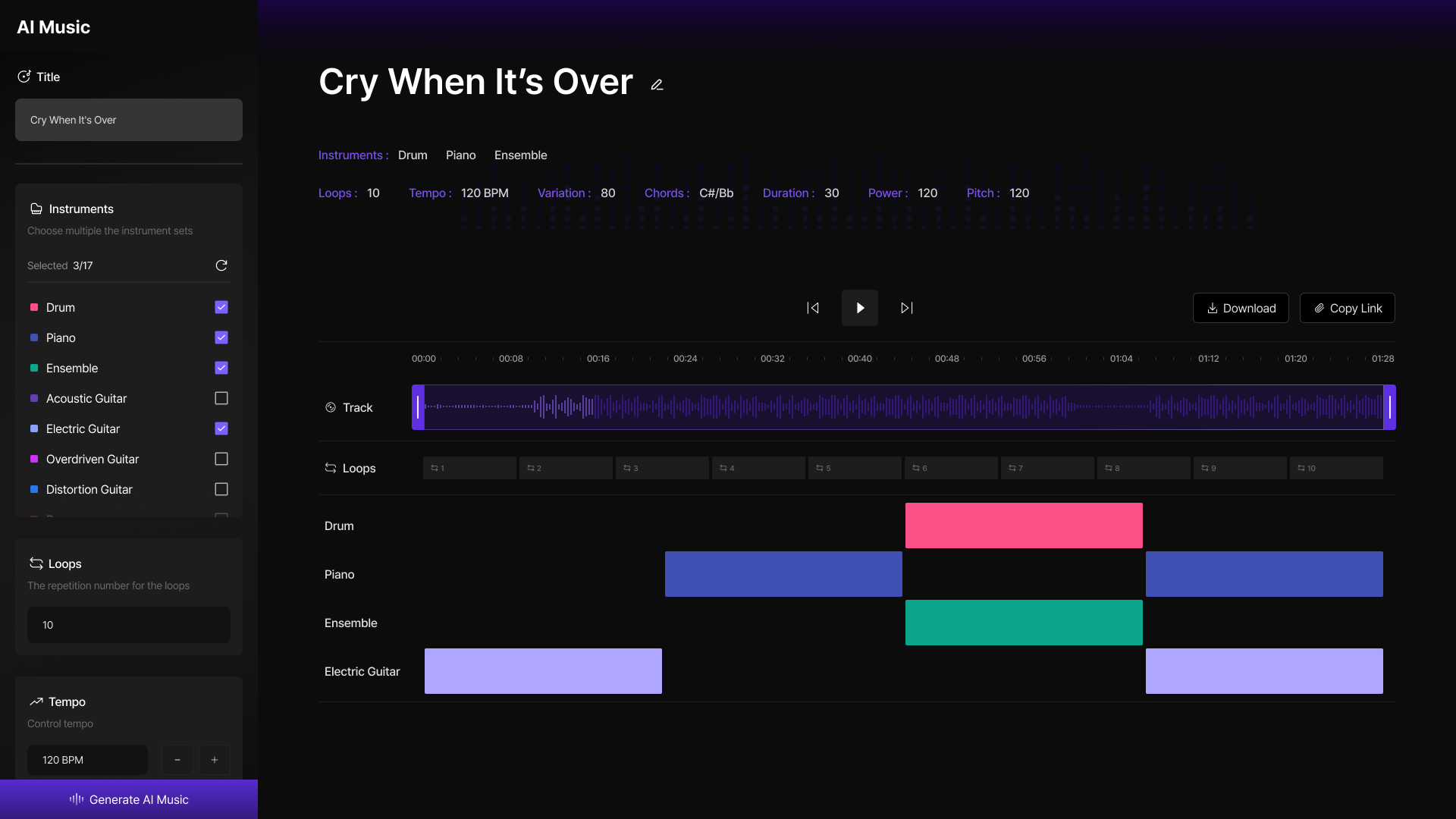} \\
  \caption{The user interface of our demonstration.}
\end{figure*}

Our demonstration consists of a sidebar and a central interactive panel, offering users the capability to generate and edit MIDI (Figure 1).
In the sidebar, you can specify musical options and generate 4 bars of multitrack MIDI composition.
In addition, you can adjust the number of repetitions of generated samples and control sampling diversity (\textit{e.g.} temperature).
In the central panel, following the generation, the generated music can be edited by either adding/removing instruments or adjusting playtime.

\section{Methods}

In this section, we introduce the process for data preparation and the proposed autoregressive conditional generation.

\subsection{Data Preparation}

To realize multi-track symbolic music generation, we choose datasets containing a diverse range of musical genres, styles, and instruments, which are Lakh MIDI Dataset (LMD)~\cite{raffel2016learning} and MetaMIDI Dataset (MMD)~\cite{ens2021building}.
LMD is widely used since it comprises 176,581 MIDI files spanning diverse genres and tracks.
Also, MMD can enhance our limited datasets since they have collected large-scale MIDIs up to 436,631 files.
Additionally, we make use of EMOPIA~\cite{hung2021emopia} and POP1k7~\cite{hsiao2021compound} for the training and test.
We convert each MIDI into REMI+ representation~\cite{von2022figaro}, an extended version of REMI~\cite{huang2020pop} that enables the expression of multiple tracks.
Our vocabulary contains 528 events for 9 categories (1 〈bar〉, 32 〈tempo〉, 17 〈instrument〉, 128 〈pitch〉, 128 〈pitch drum〉, 48 〈position〉, 58 〈duration〉, 32 〈velocity〉, and 84 〈chord〉) and several special tokens.
We adopt the same configuration for REMI+ as described in ~\cite{von2022figaro}, eliminating non-4/4 signature music.
Since the REMI+ serves as a universal MIDI representation for multitrack music, we restrict the music length to 4 bars due to computational constraints.

\subsection{Music Generation}

Our music generation model is designed to incorporate music-related metadata as input conditions with a decoder-only autoregressive Transformer.
The metadata for our system encompasses a combination of instrument set, mean pitch, mean tempo, mean velocity, mean duration, and chord set.
As shown in the upper part of Figure 2, all the conditions and music tokens participate in the next token prediction at the training phase.
However, this process prohibits the user's creativity and controllability in that 1) there are few training samples satisfying all provided conditions, and 2) non-experts may struggle to complete all conditions.

Here, during training, we apply random drops of the condition tokens for flexible controllability (the bottom part of Figure 2).
It ensures that users are not required to complete all musical conditions to generate music sequences.
While this technique is generally used to regularize or reduce the computation of Transformer~\cite{rao2021dynamicvit,koutini2021efficient}, our intention is to relax the conditions from ``and" to ``or" relationship.

\begin{figure}[!t]
  \centering
  \includegraphics[width=\linewidth]{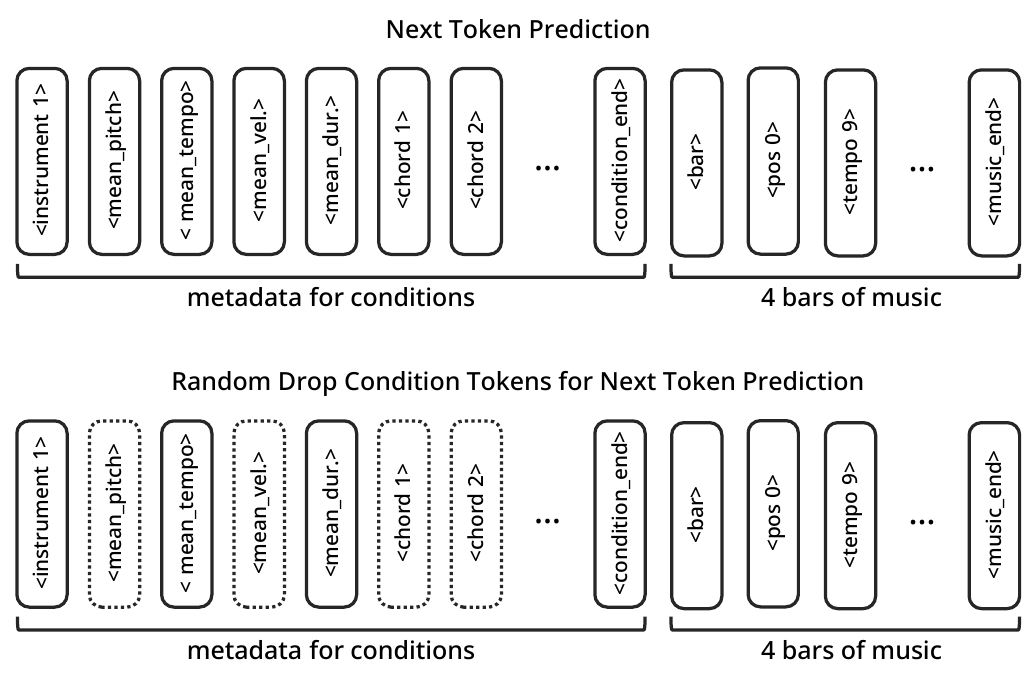} \\
  \caption{Upper: The original next token predictions, Bottom: The next token predictions with random drop conditions. The dotted boxes indicate that tokens are dropped.}
\end{figure}

\begin{table*}[!t]
    \centering
    \resizebox{\textwidth}{!}
    {
        \begin{tabular}{llrrrrrrrrrr}
            \toprule
            Input Set  & Drop  & Perplexity (\(\downarrow\))  & Density (\(\uparrow\))  & Coverage (\(\uparrow\))  & \multicolumn{7}{c}{Controllability} \\
        & & & & & I (↑) & MP (↓) & MT (↓) & MV (↓) & MD (↓) & SC (↑) & RC (↑) \\
            \midrule 
            Superset  & X  & 2.005  & 0.482  & 0.555  & 0.981  & 1.381  & 0.000  & 2.763  & 2.530  & 0.551  & 0.620 \\
            Superset  & O  & 2.034  & 0.481  & 0.466  & 0.971  & 2.018  &  0.018 &  3.879 &  2.891 & 0.495  & 0.579 \\
            Subset  & X  & 2.309  & 0.214  & 0.338  & 0.970  & 2.850  & 0.000  & 2.845  &  2.677  & 0.369  & 0.415  \\
            Subset  & O  & 2.026  & 0.499  & 0.374  & 0.970  & 1.989  & 0.000  & 4.004  & 2.955  & 0.513  & 0.602 \\
            \bottomrule
        \end{tabular}
    }
    \caption{The evaluation table of music generation conditioned on musical metadata.
    Every sample is generated in a greedy search manner, taking the most probable token for predicting the next token.}
    \label{tab:booktabs}
\end{table*}

\subsection{Model and Training Descriptions}

Our generation model follows LLaMA changes~\cite{touvron2023llama}; pre-normalization with RMSNorm~\cite{zhang2019root}, SwiGLU activations~\cite{shazeer2020glu}, and Rotary Embeddings at each layer~\cite{su2021roformer}.
We conduct training for the 88M model scale ($n_{layers}$=12, $d_{model}$=768, $n_{heads}$=12, $d_{head}$=64) with a batch size of 256 for about 55k steps.
The AdamW~\cite{loshchilov2017decoupled} optimizer is regulated with a warm-up period of 2,000 steps and linear decay.
The training of the model involves utilizing LMD, MMD, EMOPIA, and POP1k7 datasets, and its evaluation is conducted on their held-out-song test set (0.5\%).
To reduce the computational cost, we apply byte-pair encoding~\cite{fradet2023byte} to music tokens, resulting in a token length of approximately 66\% ($\approx3$B tokens).

\section{Results}

\subsection{Quantitative Results}

We evaluate our generative model in terms of model metric, similarity evaluation, and controllability.
First, the model metric is to evaluate the model's confidence in generating samples, which is quantified through perplexity.
Second, the similarity evaluation is to measure the overlapped ratio between true and generated samples.
We evaluate the similarity using density and coverage~\cite{naeem2020reliable} which are the variants of precision and recall to estimate sample fidelity and diversity.
Since they should be evaluated on latent space, we employ the trained model to project our true and generated samples.
The embeddings from the last Transformer block are averaged for the evaluations.
We try to balance the quantities of training and generated samples and set the nearest neighbor value $k$ of the density and coverage as 5, as reported in~\cite{naeem2020reliable}. 
Lastly, controllability is how the musical properties of generated samples correspond with the provided musical metadata.
It can be described as the Jaccard Index of instrument sets (I), the absolute difference of mean pitch, mean tempo, mean velocity, mean duration (MP, MT, MV, MD), and the accuracy of predicting chord set under strict/relaxed criteria (SC, RC).
The SC increases with accurate predictions of both chord and its quality, while the RC increases with accurate predictions of chord only.

In Table 1, we report the performance of music generation under different input conditions during the inference (Input Set) and the presence of random token drops in the training process (Drop).
To form a subset of conditions, we randomly drop metadata tokens from the superset which contains all conditions.
When providing the superset as input conditions, applying the random drops slightly degrades the performance across all metrics.
However, the effectiveness of our approach becomes notable when dealing with the subset.
Since the model without applying the random drops has never encountered a partial set of conditions, the performances are significantly degraded in terms of perplexity, density, coverage, MP, SC, and RC.
In that setting, despite marginal improvement in some controllability metrics, 0.9\% of generated samples exhibit severe syntax error (\textit{e.g.} missing targeted value), so they are excluded from the results.
Compared to that, our proposed approach, incorporating the random drop, can provide selective options based on musical knowledge with minimal performance compromise.

\subsection{Human Listening Test}

\begin{figure}[!t]
  \centering
  \includegraphics[width=\linewidth]{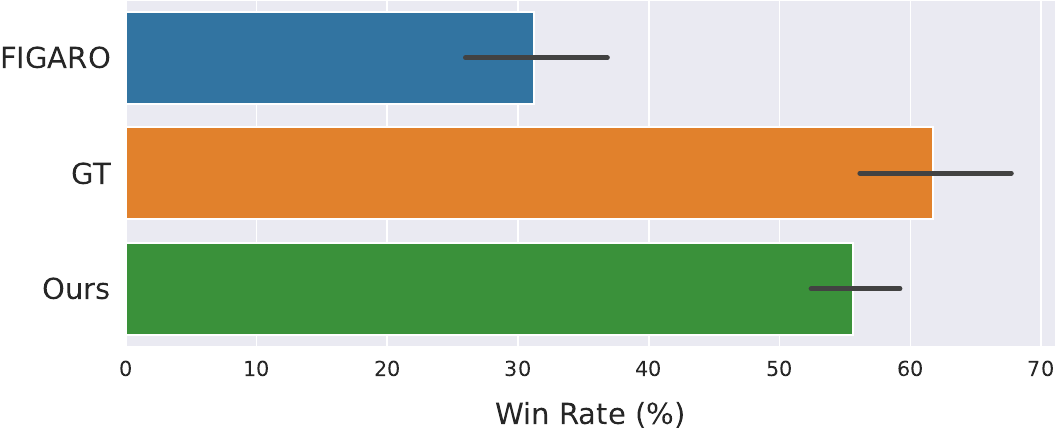} \\
  \caption{Win rates of our generated samples compared to the ground truth (GT) and FIGARO. The solid line indicates the standard deviation.}
\end{figure}

In this human listening test, we increase the model's scale to 306M ($n_{layers}$=24, $d_{model}$=1024, $n_{heads}$=16, $d_{head}$=64) and compare the nuanced quality of generated samples with FIGARO$_{expert}$~\cite{von2022figaro} and training set (ground truth).
Our model is trained with randomly dropping conditions over 58k steps with a batch size of 512.
The model generates music samples, conditioned on the subset of musical metadata for the held-out-song test set (under top-$k$ sampling ($k=5$) at a temperature of 1).
FIGARO$_{expert}$ serves as a good baseline since it relies solely on human-interpretable metadata, similar to our approach, and generates samples with control at the bar-level.\footnote{We modify the official implementation of FIGARO, which is available on Google Colab (\url{https://tinyurl.com/28etxz27}).}
It also generates samples conditioned on the identical set as ours, but it requires the superset of conditions along with note density (chosen randomly from 1 to 3).

We have assigned 20 sets of pairwise comparisons randomly sampled from three groups to each of the 30 individuals for evaluation.\footnote{You can listen to our samples on \url{https://tinyurl.com/nhjp2jww}.}
They have been asked to indicate their preference between two different samples.
As Figure 3, our proposed model surpasses FIGARO$_{expert}$ and closely follows the performance of the ground truth.
Although there is room to explore different configurations of FIGARO$_{expert}$ such as note density and temperature, this result demonstrates the effectiveness of our model in music quality.

\section{Conclusion}

We have presented the demonstration of symbolic music generation, which generates 4 bars of multitrack MIDI with flexible control.
The quantitative results demonstrate that users can selectively combine input conditions to generate plausible music samples.
Moreover, the human listening test verifies that our model can generate samples that closely resemble human-created music.
Our model, trained to generate 4 bars of music with global control, has limitations in extending music length and controlling bar-level local elements.
However, our attempts hold significance in generating high-quality musical themes that can be used as loops.

\appendix

\section*{Ethical Statement}
It is essential to prioritize the responsible development and use of AI in music generation to ensure the integrity, authenticity, and fair representation of creative works, acknowledging potential limitations and considering the impact on artistic expression and cultural values.

\bibliographystyle{named}
\bibliography{ijcai24}

\end{document}